\documentclass[nofootinbib,prd,a4paper,showpacs]{revtex4} 
\usepackage{amssymb,amsmath,epsfig}

\usepackage{color, ulem}

\usepackage{epsfig}

\def\nn{\nonumber}        

\newcommand{\bm}[1]{\mbox{\boldmath $#1$}}

\newcommand{\open}{{<\kern -0.3 em{\scriptscriptstyle )}}}

\newcommand{\nslash}{\kern 0.2 em n\kern -0.45em /}
\newcommand{\Pslash}{\kern 0.2 em P\kern -0.56em \raisebox{0.3ex}{/}}
\newcommand{\pslash}{\kern 0.2 em p\kern -0.4em /}
\newcommand{\kslash}{\kern 0.2 em k\kern -0.45em /}
\newcommand{\Sslash}{\kern 0.2 em S\kern -0.56em \raisebox{0.3ex}{/}}

\newcommand{\eq}{\begin{equation}}
\newcommand{\ee}{\end{equation}}
\newcommand{\beq}{\begin{equation}}
\newcommand{\eeq}{\end{equation}}
\newcommand{\ba}{\begin{eqnarray}}
\newcommand{\ea}{\end{eqnarray}}
\newcommand{\eqa}{\begin{eqnarray}}
\newcommand{\eea}{\end{eqnarray}}

\newcommand{\sumint}{\kern 0.2 em {\textstyle\sum} \kern -1.1 em \int}

\newcommand{\simorder}{\raisebox{-4pt}{$\, \stackrel{\textstyle >}{\sim} \,$}}
\newcommand{\simorderr}{\raisebox{-4pt}{$\, \stackrel{\textstyle <}{\sim} \,$}}

\DeclareMathOperator{\tr}{Tr}

\begin{document} 

\title{TMD evolution and the Higgs transverse momentum distribution}

\author{Dani\"el Boer}
\email{d.boer@rug.nl}
\affiliation{University of Groningen, Nijenborgh 4, NL-9747 AG Groningen, The Netherlands}

\author{Wilco J. den Dunnen}
\email{wilco.den-dunnen@uni-tuebingen.de}
\affiliation{Institute for Theoretical Physics,
                Universit\"{a}t T\"{u}bingen,
                Auf der Morgenstelle 14,
                D-72076 T\"{u}bingen, Germany}
                
\date{\today}

\begin{abstract}
The effect of the linear polarization of gluons on the transverse momentum distribution in Higgs production is studied within the framework of TMD factorization. 
For this purpose we consider the TMD evolution for general colorless scalar boson production, from the lower mass
$C$-even scalar quarkonium states $\chi_{c0}$ and $\chi_{b0}$ to the Higgs mass scale. In the absence of an intrinsic
nonperturbative linearly polarized gluon distribution the results correspond to the CSS formalism, indicating a rather
rapid decrease with increasing energy scale. At the Higgs mass scale the contribution from linearly polarized gluons is
in this case found to be on the percent level, somewhat larger than an earlier finding in the literature. At the
lower mass scale of quarkonium states $\chi_{c0}$ and $\chi_{b0}$ we find contributions at the 15-70\% level, albeit with
considerable uncertainty. In the presence of an intrinsic linear gluon polarization, percent level effects are also found at
the Higgs mass scale, but with a considerably slower evolution. Although these results were obtained using a model for
the TMDs that are approximately Gaussian at small transverse momenta 
and have the correct perturbative power law fall off at large transverse momenta, it illustrates well the differences that can exist 
between results obtained from a TMD formalism as compared to a CSS formalism. The behavior of the TMDs at small $p_T$ can affect
the results for all transverse momenta of the produced boson, even for a particle as heavy as the Higgs. The TMD evolution from $\chi_{c0}$
to $\chi_{b0}$ may be used to constrain the nonperturbative contributions and improve on the prediction of the effect at
the Higgs mass scale.
\end{abstract}


\maketitle


\section{Introduction}
It is known that gluons inside unpolarized hadrons in principle can have nonzero linear polarization \cite{Mulders:2000sh}. 
This requires nonzero transverse momentum of the gluons and therefore, the distribution describing the amount of polarization is a 
transverse momentum dependent distribution function (TMD). In Refs.\ \cite{Sun:2011iw,Boer:2011kf} it was discussed how 
this linear polarization affects the Higgs transverse momentum distribution. The presence of linear polarization of gluons also 
became apparent from the calculation of perturbative corrections to gluon-gluon scattering processes, including Higgs production 
\cite{Nadolsky:2007ba,Catani:2010pd,Catani:2011kr,Wang:2012xs,Becher:2012yn}. 
Linear polarization of gluons is perturbatively generated at order $\alpha_s$, entering  
with new coefficient functions $G$ ($I^{(2)}$ in \cite{Becher:2012yn}), 
which are driven by the ordinary collinear gluon and quark distribution functions. 
It means that if a linearly polarized gluon distribution function is not intrinsically present nonperturbatively, it will 
in any case be generated perturbatively. In general, one expects both perturbative and nonperturbative contributions, with relative 
magnitudes that depend on the energy scale. 

The transverse momentum distribution of Higgs bosons produced in gluon-gluon fusion is sensitive to the 
polarization of the gluons. This TMD observable was suggested as a new way to determine whether the Higgs boson is a scalar or 
pseudoscalar boson \cite{Boer:2011kf}, an issue essentially settled by now, but it can also be used to probe anomalous couplings of 
the Higgs boson more generally \cite{Boer:2013fca}. As the observed Higgs boson has a mass of about 126 GeV, the effect of 
QCD corrections is expected to be significant w.r.t.\ the tree level analyses of Refs.\ \cite{Boer:2011kf,Boer:2013fca}. 
It is the goal here to investigate this observable beyond tree level within the TMD approach. 

Although not explicitly shown to all orders yet, TMD factorization is expected to hold in Higgs production.
TMD factorization theorems have been established for various processes, such as the Drell-Yan process and 
semi-inclusive DIS \cite{CS81,JMY,Collins:2011zzd,Collins:2011ca,GarciaEchevarria:2011rb}. 
The corresponding evolution equations of the TMDs are known (at least) to order $\alpha_s$
and yield the leading order scale dependence of the cross section. Inclusion of polarization of initial or final state hadrons 
and/or of the partons involved leads to asymmetries in the cross section expression, such as the  
Sivers and Collins asymmetries. The TMD evolution of such asymmetries has been studied in e.g.\
\cite{B01,B09,AR,ACQR, APR, Anselmino:2012aa,Sun:2013dya,Boer:2013zca,Echevarria:2014xaa,Echevarria:2014rua}. 

Assuming TMD factorization to hold in Higgs production, the energy scale dependence will correspondingly be dictated by 
TMD evolution. Although one cannot physically change the Higgs mass 
$m_H$, one can nevertheless study how the observable would change in the production of a colorless scalar boson with varying 
mass $Q$. In this way one can draw conclusions about the size and shape of the distribution at the actual Higgs mass $Q=m_H$. 
Moreover, the calculation will be applied to the production of $C$-even scalar quarkonium states $\chi_{c0}$ and $\chi_{b0}$ 
at $Q=3.4$ GeV and $Q=9.9$ GeV, respectively \cite{Boer:2012bt}.  
The aim of this paper is to study this TMD evolution and its consequences. 

In the TMD formalism the differential cross section of colorless scalar boson production at small transverse momentum 
($\bm{q}_T^2 \equiv Q_T^2 \ll Q^2$) is written as
\beq
\frac{d\sigma}{dx_A dx_B d\Omega {d^2 \bm{q}_T^{}} } = \int d^2 {b} 
\, e^{-i {\bm{b} \cdot \bm{q}_T^{}}} \tilde{W}({\bm{b}}, Q; x_A, x_B) + {\cal O}\left(\frac{Q_T^2}{Q^2}\right), 
\label{diffxs}
\eeq
where $x_{A(B)} = Q^2/(2P_{A(B)}\cdot q)$ are the observed Bjorken variables for hadron momenta $P_A$ and $P_B$ and 
momentum $q$ of the produced scalar boson, which sets the energy scale $Q$ through $q^2=Q^2$. Furthermore, 
$Q_T$ indicates the scalar boson's momentum transverse to the beam axis in the hadron center of mass frame.
Here we will focus on the angular averaged case, so angle definitions will not matter. 
Upon ignoring possible gluon polarization effects, the integrand $\tilde{W}({\bm{b}}, Q; x_A, x_B)$ for the 
gluon-gluon fusion process will be   
\ba
\tilde{W}({\bm{b}}, Q; x_A, x_B) & = & \tilde{f}_1^{g}(x_A,{\bm{b}^2};\zeta_A, \mu)
\, \tilde{f}_1^{g}(x_B,{\bm{b}^2};\zeta_B,\mu) H\left(Q;\mu\right),
\label{Wtilde}
\ea
which is an expression in terms of the Fourier transforms $\tilde{f}_1^g(x,\bm{b}^2)$ of the
unpolarized gluon TMDs $f_1^g(x,\bm{k}_T^2)$ and the partonic hard scattering factor $H$, 
which for the choice $\mu = Q$ just becomes a finite expression in terms of $\alpha_s(Q)$.
Taking $\mu=Q$ in the TMDs will lead to large logarithmic terms for small $b$ values that require resummation. 
Running down the TMDs to a lower ($b$-dependent) scale removes these large logarithms from the TMDs
and resums them in the form of a Sudakov factor.  
The dependence of the TMDs on $\zeta_{A(B)}$ and $\mu$ will be discussed in more detail below. 
Once the factorization expression is given, with all its scale dependences, the evolution of 
TMD cross sections follows automatically. Note that a soft factor, which arises due to soft gluon radiation, is in principle present too, 
but can be factorized and included in the definition of the TMDs \cite{Collins:2011zzd,Echevarria:2012js}. 

For large $Q_T$ the result is dominated by small $b$ values where the $b$ dependence of the TMDs can be calculated 
perturbatively and upon insertion in Eq.\ (\ref{Wtilde}) can directly be translated into the resummed CSS expression for this
process. It is important to note that only in this limit of large $Q_T$ (or equivalently, small $b$), the differential cross 
section expression will involve integrals over the {\it partonic} momentum fractions. 
Allowing for polarization of gluons to be present or to develop under evolution, forces the inclusion of another 
gluon TMD, here denoted by $h_1^{\perp g}$, or of the above-mentioned $G$ coefficient functions.  
In the recent study \cite{Wang:2012xs} the effect 
of the $G$ functions on the resummed transverse-momentum distribution in the gluon fusion process is found to be below the 
percent level. In that reference the CSS expression is studied, 
derived from resummation of the large $Q_T$ fixed order calculation within collinear factorization,
whereas here the TMD factorized expression including a possible 
nonperturbative linearly polarized gluon distribution will be studied. 
It turns out that such a nonperturbative contribution can significantly modify the results, especially the evolution.
Since this intrinsic nonperturbative contribution cannot be calculated, the results come with a considerable uncertainty
though.
 
Finally, we note that the linear polarization of gluons inside unpolarized hadrons enters observables other than 
Higgs production, so even if the effect turns out to be small, below or possibly at the limit of what is observable, it can 
nevertheless be studied in other ways, cf.\ \cite{Boer:2009nc,Boer:2010zf,Qiu:2011ai,Pisano:2013cya,Dunnen:2014eta}. 
But here we will restrict to colorless scalar boson production in proton-proton collisions, which may also apply to 
some extent to proton-nucleus collisions \cite{Schafer:2012yx}. 

\section{The Higgs transverse momentum distribution at tree level}

In collinear factorization the transverse momentum distribution of Higgs production at tree level would be proportional to 
a delta function at zero transverse momentum. This is of course an unrealistic approximation to the real distribution that is 
affected by radiative corrections and by the transverse momentum distribution of gluons inside the colliding protons. 
Both give rise to a contribution from linearly polarized gluons, even though the protons themselves are unpolarized.
In this section we briefly summarize the contribution of the linear polarization of gluons on the tree level distribution.  
In Ref.\ \cite{Boer:2011kf} a tree level expression for the cross section was presented: 
\begin{eqnarray}
\frac{E\,d\sigma^{p p \to H X}}{d^{3}\vec{q}}\Big|_{q_{T}\ll m_{H}} & = & \frac{\pi \sqrt{2}G_F}{128m_H^2 s}\left(\frac{\alpha_{s}}{4\pi}\right)^{2}\,\left|{\cal A}_{H}(\tau)\right|^2 \left(\mathcal{C}\left[f_{1}^{g}\, f_{1}^{g}\right]+\mathcal{C}\left[w_{H}\, h_{1}^{\perp g}\, h_{1}^{\perp g}\right]\right)\,+\mathcal{O}\left(\frac{q_{T}}{m_{H}}\right)\,,\label{Higgsdiffxs} 
 \end{eqnarray}
which involves the standard tree level TMD convolution
\begin{eqnarray}
\mathcal{C}[w\, f\, f] & \equiv & \int d^{2}\bm p_{T}\int d^{2}\bm k_{T}\,
\delta^{2}(\bm p_{T}+\bm k_{T}-\bm q_{T})\, w(\bm p_{T},\bm k_{T})\, f(x_{A},\bm p_{T}^{2})\, f(x_{B},\bm k_{T}^{2})\,,\label{eq:Conv}
\end{eqnarray} 
with the weight $w_H$ defined as:
\begin{equation}
w_{H}=\frac{(\bm p_{T}\cdot\bm k_{T})^{2}-\tfrac{1}{2}\bm p_{T}^{2}
\bm k_{T}^{2}}{2M^4}\,.\label{Higgsweight}
\end{equation}
Here $s=(P_A+ P_B)^2$ is the center of mass energy squared, $M$ denotes the proton mass, and ${\cal A}_{H}(\tau)$ is 
a function of $\tau= m_H^2/(4m_t^2)$ with $m_t$ the top quark mass, the explicit expression of which will not be needed here. 

The transverse momentum dependent distribution functions (TMDs) of gluons inside an unpolarized proton are defined through a 
correlator of gluon field strengths \cite{Mulders:2000sh} which (omitting gauge links) is given by  
\begin{eqnarray}
\Phi_g^{\mu\nu}(x,\bm p_T )
& = &  \frac{n_\rho\,n_\sigma}{(p{\cdot}n)^2}
{\int}\frac{d(\xi{\cdot}P)\,d^2\xi_T}{(2\pi)^3}\
e^{ip\cdot\xi}\, \langle P|\,\tr\big[\,F^{\mu\rho}(0)\,
F^{\nu\sigma}(\xi)\,\big]
\,|P \rangle\,\big\rfloor_{\text{LF}} \nonumber \\
&=&
-\frac{1}{2x}\,\bigg \{g_T^{\mu\nu}\,f_1^g
-\bigg(\frac{p_T^\mu p_T^\nu}{M^2}\,
{+}\,g_T^{\mu\nu}\frac{\bm p_T^2}{2M^2}\bigg)
\;h_1^{\perp\,g} \bigg \} ,\label{GluonCorr}
\end{eqnarray}
with $p_{T}^2 = -\bm p_{T}^2$, $g^{\mu\nu}_{T} = g^{\mu\nu}
- P^{\mu}n^{\nu}/P{\cdot}n-n^{\mu}P^{\nu}/P{\cdot}n$.
Here the gluon momentum is decomposed as $p = x\,P + p_T + p^- n$, with $n$ a lightlike vector
conjugate to the parent hadron's four-momentum $P$. 
The two gluon TMDs, $f_1^g(x,\bm{p}_T^2)$ and $h_1^{\perp\,g}(x,\bm{p}_T^2)$, represent the  
unpolarized and linearly polarized gluon distributions, respectively. 
 
The presence of linearly polarized gluons does not affect the transverse momentum integrated 
Higgs production cross section, as can be explicitly verified by integrating the expression in 
Eq.\ (\ref{Higgsdiffxs}) over ${\bm q}_T$. It can also be seen that the integration weighted with an 
additional factor of $q_T^2$ vanishes, i.e.\ $\int d^{2}\bm q_{T}\, \bm q_T^{2}\,
\mathcal{C}[w_{H}\, h_{1}^{\perp g}\, h_{1}^{\perp g}]=0$. This implies that the transverse 
momentum distribution of the $h_1 ^{\perp g}$ dependent term exhibits a double node 
in $\bm q_T$, unless the neglected $Q_T^2/Q^2$ contributions modify this behavior. 
Any node in the $h_1 ^{\perp g}$ contribution will show up as a modulation on top of the larger  
contribution from unpolarized gluons, leading to a total transverse momentum distribution of 
Higgs bosons that must be positive definite. It is the goal of this paper to investigate the 
energy scale dependence of this modulation in order to get a better idea about its expected 
shape and magnitude at the Higgs mass scale.  
In other words, we wish to study the scale dependence of the dimensionless ratio 
\beq
{\cal R}(Q_T) \equiv \frac{\mathcal{C}[w_H\,h_1^{\perp g}\,h_1^{\perp g}]}{\mathcal{C}[f^g_1\,f_1^g]}.
\eeq 
As the Higgs boson mass is 
around 126 GeV, the effect of higher order corrections is expected to be significant.

\section{The Higgs transverse momentum distribution beyond tree level}
Beyond tree level, TMDs are not only functions of a momentum fraction $x$ and the transverse momentum 
$p_T$, but also will depend on a renormalization scale $\mu$. In order to avoid large logarithmic terms in the hard scattering 
or in the TMDs, the renormalization scale will be chosen as $\mu=Q$ in the hard scattering, such that $H \propto 1
+\alpha_s \times \text{finite}$, and the TMDs will be evolved from the high scale $Q$ to the scale $\mu_b=b_0/b=2e^{-\gamma_E}/b$ 
($b_0 \approx 1.123$), where $b$ is the Fourier conjugate of the transverse momentum. 
We will use the simplified notation $\bm{b}=\bm{b}_T$ and $b^2=\bm{b}^2$, which is not to be confused with $b_\mu b^\mu$. 

Evolving the TMDs down from the scale $Q$ to $\mu_b$ introduces, as is well-known, a Sudakov factor, here denoted by $S_A$, which enters in an 
exponential. Apart from the dependence on $\mu$, the TMDs also depend on a rapidity cut-off $\zeta_{A(B)}$, defined as:
\beq
\zeta_A = M_{P_A}^2 x_A^2 e^{2(y_{A}-y_n)},\quad 
\zeta_B = M_{P_B}^2 x_B^2 e^{2(y_n-y_{B})},
\eeq
where $y_{A(B)}$ denotes the rapidity of hadron $A(B)$. The dependence on the arbitrary rapidity cut-off $y_n$ cancels in the cross section, 
which only depends on the combination $\zeta_A \zeta_B \approx Q^4$. 

The evolution in $\zeta$ and $\mu$ is given by the Collins-Soper and Renormalization Group equations, respectively \cite{Collins:2011zzd,AR}. 
With these evolution equations one can evolve the TMDs down to the scale $\mu_b$. 
One could also consider evolving the TMDs down to a fixed scale $Q_0$ inside the range 
of validity of perturbation theory. However, for the relatively large transverse momenta
to be considered here this does not seem an appropriate choice, because unresummed logarithms of $b^2 Q_0^2$ can become large. 

The above tree level formula for the $f_1^g$ term will upon inclusion of $\alpha_s$ corrections become:
\ba 
\mathcal{C}\left[f_{1}^{g}\, f_{1}^{g}\right]& = & \; \int 
\frac{d^2 \bm{b}}{(2\pi)^2} \, e^{i \bm{b} \cdot \bm{q}_T^{}} 
\, \tilde{f}_1^g(x_{A},b^2; \zeta_A, \mu) \, 
\tilde{f}_1^g(x_{B},b^2;\zeta_B, \mu) \nn \\
& = & \; \int 
\frac{d^2 \bm{b}}{(2\pi)^2} \, e^{i \bm{b} \cdot \bm{q}_T^{}} 
\, e^{-S_A(b,Q)} \tilde{f}_1^g(x_{A},b^2; \mu_b^2, \mu_b) \, 
\tilde{f}_1^g(x_{B},b^2;\mu_b^2, \mu_b) , 
\label{conv2}
\ea
with the following {\it perturbative} Sudakov factor \cite{Catani:1988vd,Kauffman:1991cx,Yuan:1991we}:
\beq
S_A(b,Q) = \frac{C_A}{\pi} \int_{\mu_b^2}^{Q^2} \frac{d\mu^2}{\mu^2} \alpha_s(\mu)\left[\ln\left(\frac{Q^2}{\mu^2}\right)- \frac{11-2n_f/C_A}{6}\right] + {\cal O}(\alpha_s^2). \label{SAmub}
\eeq
Next-to-next-to-leading logarithmic corrections are known too \cite{deFlorian:2000pr,deFlorian:2001zd} and their effect on the resummed transverse momentum 
distribution of the Higgs boson has been studied in detail in \cite{Balazs:2000wv,Bozzi:2005wk,Bozzi:2007pn,deFlorian:2011xf,deFlorian:2012mx}.
Including the one-loop running of $\alpha_s$ one can perform the $\mu$ integral explicitly:
\beq
S_A(b,Q) = -\frac{36}{33-2n_f} \left[ \ln\left(\frac{Q^2}{\mu_b^2}\right)+
\ln\left(\frac{Q^2}{\Lambda^2}\right)\; \ln\left(1- \frac{\ln\left( 
Q^2/\mu_b^2\right)}{\ln\left(Q^2/\Lambda^2\right)} \right) + \frac{11-2n_f/C_A}{6} \ln\left(\frac{\ln\left( 
Q^2/\Lambda^2\right)}{\ln\left(\mu_b^2/\Lambda^2\right)} \right)\right]. \label{SAmubeval}
\eeq

The above expressions for the Sudakov factor are valid in the perturbative region $b\ll \Lambda_{\rm QCD}^{-1}$.
However, using the perturbative Sudakov factor at small transverse momenta is not appropriate. Inclusion of a nonperturbative Sudakov factor is necessary. 
We will follow the $b_*$ method \cite{CSS-85}, in which one replace $b \to b_*=b/\sqrt{1+b^2/b_{\max}^2}$, such that $b_*$ is always smaller
than $b_{\max}$. One then rewrites the $b$-integrand $\tilde{W}(b)$ in the standard way \cite{CSS-85} as:
\beq
\tilde{W}(b) \equiv \tilde{W}(b_*) \, e^{-S_{NP}(b)} ,
\eeq
such that for $\tilde{W}(b_*)$ the perturbative expression {\it is} valid. Here we will use the recent nonperturbative Sudakov factor $S_{NP}$ by Aybat and Rogers \cite{AR}:
\beq
S_{NP}(b,Q) = \left[g_2 
\ln\frac{Q}{2Q_0} +g_1\left(1+2g_3 \ln \frac{10xx_0}{x_0+x}\right)\right] b^2 ,
\label{ARSNP}
\eeq
with $g_1=0.201\,\text{GeV}^{2}, g_2=0.184\,\text{GeV}^{2}, g_3=-0.129, x_0=0.009,  
Q_0=1.6 \,\text{GeV}$ and $b_{\max}= 1.5 \,\text{GeV}^{-1}$. One reason for selecting this parameterization is that it is constructed and fitted such that it describes low energy semi-inclusive DIS data as well as higher energy Drell-Yan and $Z$ boson production data. Another reason is that it employs $b_{\max}= 1.5 \,\text{GeV}^{-1}$, which is favored both theoretically \cite{Echevarria:2013aca} as well as experimentally \cite{Konychev:2005iy}.
Although this $S_{NP}$ is $x$ dependent, here $x=0.09$ is chosen for simplicity (like in the numerical studies of \cite{AR,Boer:2013zca}), leading to a Gaussian with a $Q$-dependent width:
\beq
S_{NP}(b,Q) = \left[0.184 \ln\frac{Q}{2Q_0}+0.332 \right] b^2 .
\label{actualSNP}
\eeq
Including the $x$ dependence hardly affects the results at $Q=m_H$ (about 0.2\% smaller) and gives only a moderate reduction up to 10\% at low $Q$ values.
Other expressions for $S_{NP}$ have been considered in e.g.\ \cite{Landry:2002ix,Konychev:2005iy,Sun:2013dya,Sun:2013hua,Su:2014wpa}. The above factor applies to quarks, therefore, to apply it to the gluon case studied here, it seems appropriate to scale it by a factor $C_A/C_F$. Especially at low $Q$ this will make a noticeable (suppressing) difference, adding to the uncertainty of the end result. 

Now we turn to the $h_1^\perp$ term. If one considers the correlator in transverse coordinate space (restricting to transverse indices only):
\begin{eqnarray}
\tilde \Phi_g^{ij}(x,\bm b )
&= & \frac{1}{2x}\,\bigg \{\delta^{ij}\,\tilde{f}_1^g(x,b^2)
-\bigg( \frac{2b^i b^j}{b^2}\,
{-}\, \delta^{ij}\bigg)
\;\tilde{h}_1^{\perp\,g}(x,b^2) \bigg \} ,
\label{GluonCorrb}
\end{eqnarray}
where (suppressing the scale dependences)
\beq
\tilde{h}_1^{\perp\,g}(x,b^2) = \int d^2\bm p_T^{}\; \frac{(\bm{b}\!\cdot \!
\bm p_T^{})^2 - \frac{1}{2}\bm{b}^{2} \bm p_T^{2}}{b^2 M^2}
\; e^{-i \bm{b} \cdot \bm{p}_T^{}}\; h_1^{\perp g}(x,p_T^2)=  -\pi
\int dp_T^2
\frac{p_T^2}{2M^2} J_2(bp_T) h_1^{\perp g}(x,p_T^2).\label{defhtilde}
\eeq
Note that this expression satisfies the property $\tilde{h}_1^{\perp g}(x,0)=0$ due to $J_2(0)=0$. 

With this definition the convolution term beyond tree level becomes
\beq
\mathcal{C}\left[\frac{(\bm p_{T}\cdot\bm k_{T})^{2}-\tfrac{1}{2}\bm p_{T}^{2}
\bm k_{T}^{2}}{2M^4}\, h_{1}^{\perp g}\, h_{1}^{\perp g}\right]  = 
 \int \frac{d^2 \bm{b}}{(2\pi)^2} \, e^{i \bm{b} \cdot \bm{q}_T^{}} e^{-S_A(b,Q)}
\; \tilde{h}_1^{\perp g}(x_{A},b^2; \mu_b^2,\mu_b) \; \tilde{h}_1^{\perp g}(x_{B},b^2; \mu_b^2,\mu_b).
\eeq
As discussed explicitly in \cite{Sun:2011iw}, the perturbative Sudakov factor (at least to the order considered here) turns out to be the same for the 
unpolarized gluon TMD $f_1^g$ as for the linearly polarized gluon TMD $h_1^{\perp g}$.  

Putting all this together leads to the following expression for the ratio ${\cal R}$:
\beq
{\cal R}(Q_T) = \frac{\int d^2 \bm{b} \, e^{i \bm{b} \cdot \bm{q}_T^{}} e^{-S_A(b_*,Q) - S_{NP}(b,Q)}
\; \tilde{h}_1^{\perp g}(x_{A},b_*^2; \mu_{b_*}^2,\mu_{b_*}) \; \tilde{h}_1^{\perp g}(x_{B},b_*^2;  \mu_{b_*}^2,\mu_{b_*})}{
\int d^2 \bm{b} \, e^{i \bm{b} \cdot \bm{q}_T^{}} \, e^{-S_A(b_*,Q)- S_{NP}(b,Q)} \tilde{f}_1^g(x_{A},b_*^2; \mu_{b_*}^2,\mu_{b_*}) \, 
\tilde{f}_1^g(x_{B},b_*^2; \mu_{b_*}^2,\mu_{b_*})}. 
\eeq 
The $Q$-dependent part of $S_{NP}$ is universal, whereas the $Q$-independent part generally is spin dependent. 
Therefore, one should actually allow for a somewhat different $S_{NP}$ in the numerator and denominator of the ratio ${\cal R}$. 
However, this difference will not be important at the high $Q^2$ values considered here. 

The remaining ingredient is how to deal with the TMDs as function of $b_*$. 
For $b \ll \Lambda_{\rm QCD}^{-1}$ one can consider purely the perturbative calculation of this $b_* \leq b_{\max}$ behavior, 
which determines the large transverse momentum tail of the TMDs. In general, the perturbative tails are of the form (cf.\ e.g.\ \cite{AR}): 
\beq
\tilde{f}_{g/P}(x,b^2;\mu,\zeta) = \sum_{i=g,q} \int_x^1 \frac{d\hat x}{\hat x} {C}_{i/g}(x/\hat x, b^2; g(\mu),\mu, \zeta) f_{i/P}(\hat x; \mu) + {\cal O} ((\Lambda_{\rm QCD} b)^a).
\eeq
At the $x$ values of relevance here, the quarks play a subdominant role, hence 
we will simplify the calculation by dropping the quark contribution. For the two gluon TMDs considered here the expressions 
to leading order in $\alpha_s$ are given by\footnote{We thank Miguel Garc\'ia Echevarr\'ia for pointing out an error in Eq.\ (\ref{h1perptail}).} \cite{Nadolsky:2007ba,Catani:2010pd,Catani:2011kr,Sun:2011iw}: 
\ba
\tilde{f}_1^g(x,b^2; \mu_b^2,\mu_b) & = & f_{g/P}(x; \mu_b) +  {\cal O} (\alpha_s),  \label{f1tail}\\ 
\tilde{h}_1^{\perp g}(x,b^2; \mu_b^2,\mu_b) & = & \frac{\alpha_s(\mu_b) C_A}{\pi}
\int_x^1 \frac{d\hat x}{\hat x} \left(\frac{\hat x}{x}-1\right) f_{g/P}(\hat x; \mu_b) 
+ {\cal O} (\alpha_s^2). \label{h1perptail}
\ea
One sees that they are both determined by the collinear unpolarized gluon distribution $f_{g/P}$, but start at different orders in $\alpha_s$. 
The perturbative expression in Eq.\ (\ref{h1perptail}) satisfies the property $\tilde{h}_1^{\perp g}(x,0)=0$, because $\alpha_{s}(\infty) =0$.

For the function $\tilde{h}_1^{\perp g}(x,b^2; \mu_b^2,\mu_b)$ the coefficient function $C$ is usually denoted by $G$. 
In Ref.\ \cite{Wang:2012xs} the effect of including the $G$ functions is found to be less than one percent. 
That result is obtained using a different $S_{NP}$ and a different collinear gluon distribution, and includes higher order effects not considered here. 
Schematically, the ratio of coefficient functions entering in ${\cal R}$ is of the form: 
\begin{multline}
\frac{G^{(1)}G^{(1)} \alpha_s^2+2G^{(1)}G^{(2)}\alpha_s^3}{C^{(0)}C^{(0)} + 2C^{(0)}C^{(1)}\alpha_s + (C^{(1)}C^{(1)}+2C^{(0)}C^{(2)})\alpha_s^2} 
\approx \\[2 mm]
\frac{G^{(1)}G^{(1)} \alpha_s^2}{C^{(0)}C^{(0)}}\left(1+ \frac{2G^{(1)}G^{(2)}}{G^{(1)}G^{(1)}}\alpha_s+{\cal O}(\alpha_s^2)\right) 
\left(1- \frac{2C^{(0)}C^{(1)}}{C^{(0)}C^{(0)}}\alpha_s + {\cal O}(\alpha_s^2) \right).
\end{multline}
In this paper we will only study the first factor ($G^{(2)}$ is not yet known). In \cite{Wang:2012xs} the second factor is also dropped, but the third one is 
included even taking into account the ${\cal O}(\alpha_s^2)$ term. However, there is no reason to assume that the second factor will be smaller in magnitude than the third factor, as both are driven by the same unpolarized collinear parton distributions. The inclusion of the third factor without the second one may thus provide an underestimate. 

Upon insertion of the leading order perturbative tails of the TMDs we obtain 
the following expression for ${\cal R}$ (applicable at sufficiently small $x$ such that quark contributions can be neglected and for $\Lambda_{\rm QCD} \ll Q_T \ll Q$): 
\begin{multline}
{\cal R}(Q_T) \approx \frac{C_A^2}{\pi^2} \left[\int db b \, J_0(bQ_T^{})\,  e^{-S_A(b_*,Q) - S_{NP}(b,Q)}\, f_{g/P}(x_A; \mu_{b_*})f_{g/P}(x_B; \mu_{b_*})\right]^{-1}\\
\times \int db b \, J_0(bQ_T^{})\,  e^{-S_A(b_*,Q) - S_{NP}(b,Q)}\, \alpha_s(\mu_{b_*})^2 
\int_{x_A}^1 \frac{d\hat x}{\hat x} \left(\frac{\hat x}{x_A}-1\right) f_{g/P}(\hat x; \mu_{b_*})\,\int_{x_B}^1 \frac{d\hat x'}{\hat x'} \left(\frac{\hat x'}{x_B}-1\right) f_{g/P}(\hat x'; \mu_{b_*}). \label{Rtail}
\end{multline}
Since only the perturbative tails are included, this expression corresponds to the one of the CSS approach in leading order. It turns out to yield percent level effects at the Higgs mass scale and $\sqrt{s}=8\, {\rm TeV}$. This is about a factor of 2-3 larger than obtained in Ref.\ \cite{Wang:2012xs}, which as mentioned includes some higher order corrections, but also uses a different $S_{NP}$ (the ``BLNY'' parameterization of \cite{Landry:2002ix}) and a different collinear gluon distribution (CTEQ6.6). Employing that same $x$-dependent $S_{NP}$  from \cite{Landry:2002ix}, which has $b_{\max}= 0.5 \,\text{GeV}^{-1}$, and the CTEQ6 LO gluon distribution function in the present analysis yields results that are about 20\% smaller at $Q=m_H$ and about 30\% smaller at low $Q$ than the results presented below. It means that the variation due to choice of $S_{NP}$ and collinear gluon distribution is not that large and less important than the other uncertainties to be discussed below.
  
\begin{figure}[htb]
\begin{center}
\includegraphics[height=7 cm]{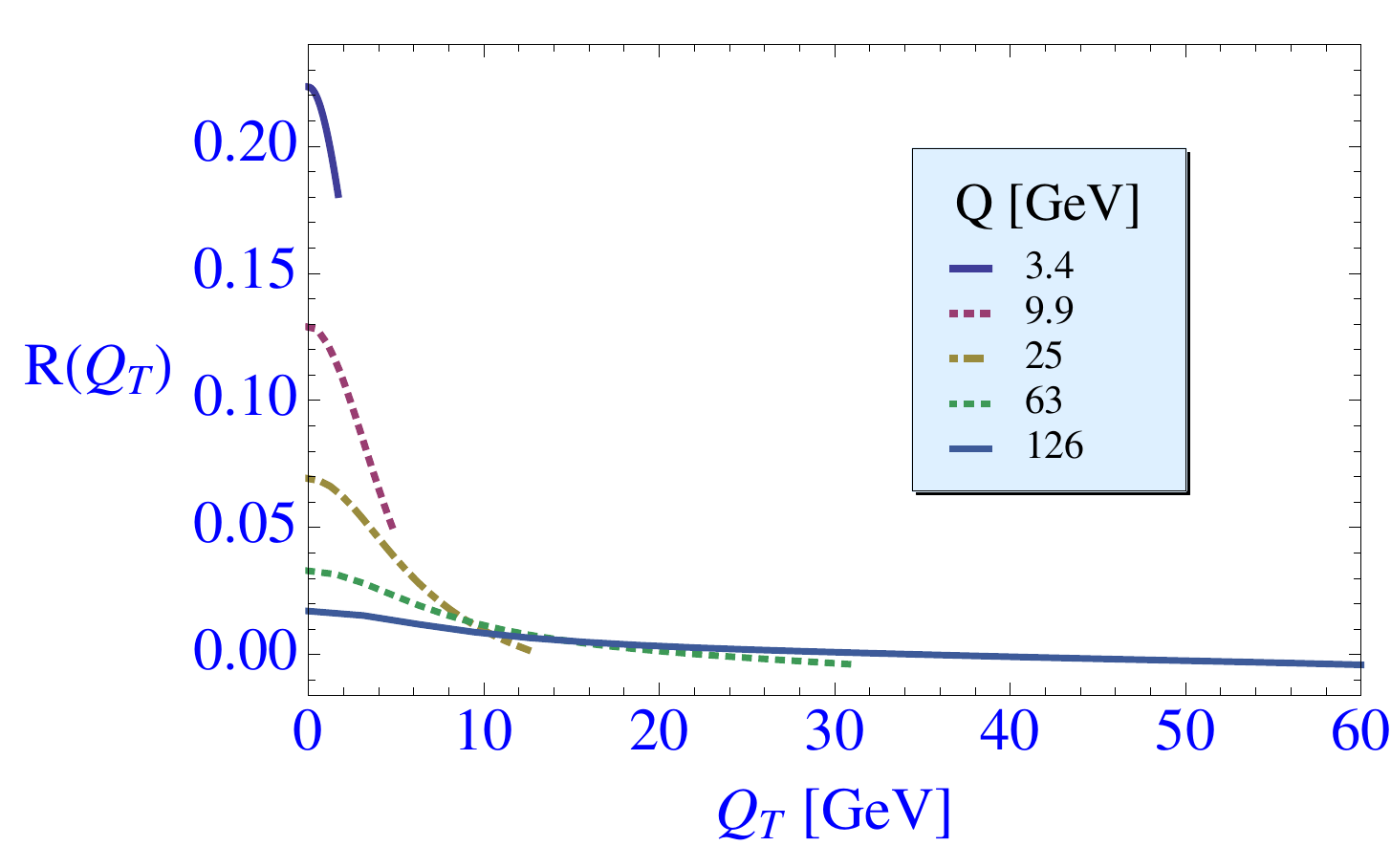}\hspace{1 cm}
\caption{${\cal R}(Q_T)$ evaluated using Eq.\ (\ref{Rtail}) at $Q=3.4, 9.9, 25, 63, 126 \, \text{GeV}$.}
\label{RplotQT}
\end{center}
\vspace{-2 mm}
\end{figure}

Numerical results are presented for $x_A=x_B=Q/(8\, {\rm TeV})$ using the leading order MSTW08 LO gluon distribution for
$n_f=5$ and $\Lambda_{\rm QCD}=0.2$ GeV for definiteness. We note that the $S_{NP}$ factor by Aybat and Rogers \cite{AR} was 
obtained using the MSTW08 parameterization.
Apart from the Higgs mass scale $Q=126$ GeV, we include
the scales $Q=3.4$ GeV (where $n_f=4$ would be more appropriate, but for simplicity we stick to fixed $n_f$) and $Q=9.9$
GeV, which correspond to the masses of the $C$-even scalar quarkonium states $\chi_{c0}$ and $\chi_{b0}$, respectively,
and some arbitrary intermediate scales. Figure \ref{RplotQT} shows the results for ${\cal R}$. 

As can be seen, it yields a ratio on the percent level in Higgs production. This may be challenging to observe at
LHC. At $Q_T=0$ the ratio falls off approximately as $1/Q^{0.85}$ for $Q \simorder 20\, {\rm GeV}$, but for lower
$Q$ values somewhat slower. At the lower mass scale of the $C$-even scalar quarkonium states $\chi_{c0}$ and
$\chi_{b0}$ we find sizeable contributions on the 10-20 percent level. The TMD evolution from $\chi_{c0}$ to $\chi_{b0}$ is about a factor
of 2 for an energy that changes by a factor of 3 approximately. This relatively fast evolution could perhaps be
observable experimentally. 

\begin{figure}[htb]
\begin{center}
\includegraphics[height=7 cm]{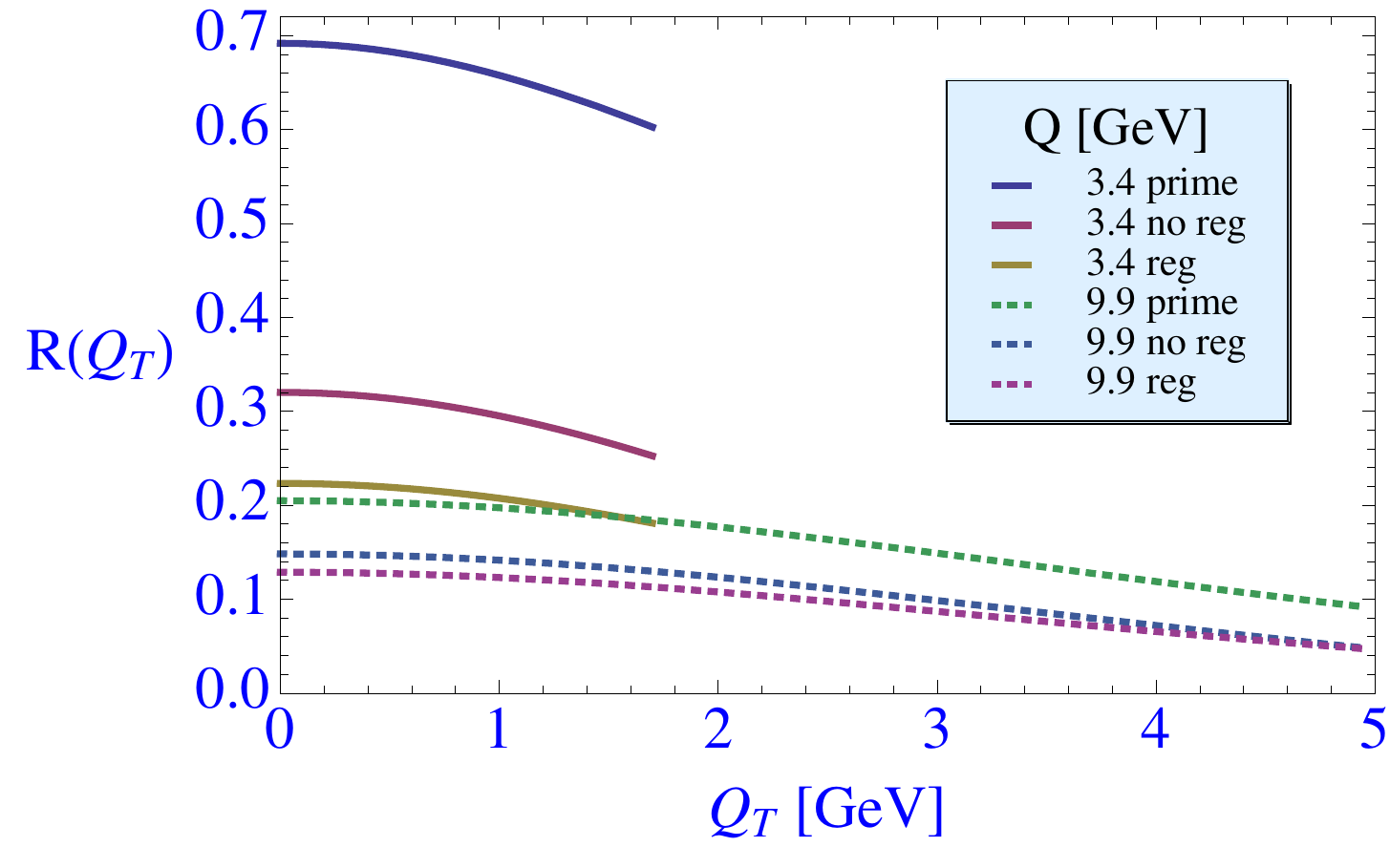}\hspace{1 cm}
\caption{${\cal R}(Q_T)$ evaluated at $Q=3.4\, \text{GeV}$ (solid lines) and $Q= 9.9\, \text{GeV}$ (dotted lines)
using $\mu_b$ (middle, ``no reg''), $\mu_b^\prime$ in $S_A$ only (lower, ``reg'') and $\mu_b^\prime$ in $S_A$ and
TMDs (upper, ``prime'').}
\label{R-regstudy}
\end{center}
\vspace{-2 mm}
\end{figure}

The $Q_T$ distributions are plotted until $Q_T \sim Q/2$ to indicate that beyond this value the
neglected $Q_T^2/Q^2$ contributions are expected to become important. Large $Q_T$ is dominated by small $b$ values, but in the region of very small $b$, i.e.\ $b \simorderr 1/Q$, perturbative expressions for $S_A$ do not have the correct behavior $S(0)= 0$. As a result the denominator in the above expression for ${\cal R}$, and consequently also the total transverse momentum distribution, does not fall off correctly at large $Q_T$. As is well-known \cite{ParisiPetronzio,CSS-85}, this requires regularization and matching onto the $Y$ term that is of order $Q_T^2/Q^2$ and neglected here. In the presented results the standard regularization $Q^2/\mu_b^2= b^2Q^2/b_0^2 \to Q^2/\mu_b^{\prime \, 2} \equiv (bQ/b_0+1)^2$ is included in $S_A$. This affects (suppresses) the result at all $Q_T$, also at $Q_T=0$ where the $Y$ term does not contribute. Although large $Q_T$ values are dominated by small $b$ values, it is important to keep in mind that all of the results are affected by the small $b$ region, no matter how 
small $Q_T$. 
The sensitivity to the regularization gives an indication of the uncertainty coming from this small-$b$ region.
Especially for lower $Q$ values the effect of regularization becomes relatively large. It results in an additional
uncertainty in the results for $\chi_{c0}$ and $\chi_{b0}$ production the size of which can be estimated by
comparing the regulated and unregulated results, and by considering different ways of regularizing the small-$b$
region. 
A different way of treating the small-$b$ region is to evolve the TMDs to the scale $\mu_b^\prime=Qb_0/(Qb+b_0)$ using: 
\beq 
\mathcal{C}\left[f_{1}^{g}\, f_{1}^{g}\right] = \; \int 
\frac{d^2 \bm{b}}{(2\pi)^2} \, e^{i \bm{b} \cdot \bm{q}_T^{}} 
\, e^{-S_A(b,Q,Q_0)} \tilde{f}_1^g(x_{A},b^2; Q_0^2, Q_0) \, 
\tilde{f}_1^g(x_{B},b^2;Q_0^2, Q_0), 
\label{conv2prime}
\eeq
with $Q_0=\mu_b^\prime$. The perturbative Sudakov factor is now (cf.\ \cite{Boer:2013zca})
\beq
S_A(b,Q,Q_0) =  - \frac{C_A}{\pi} \ln\left(\frac{Q^2}{Q_0^2}\right) \int_{Q_0^2}^{\mu_b^2} \frac{d\mu^2}{\mu^2} \alpha_s(\mu) 
+  \frac{C_A}{\pi} 
\int_{Q_0^2}^{Q^2}  \frac{d\mu^2}{\mu^2} \alpha_s(\mu)\left[\ln\left(\frac{Q^2}{\mu^2}\right)- \frac{11-2n_f/C_A}{6}\right]+ {\cal O}(\alpha_s^2).
\label{SAQ0}
\eeq
By replacing $Q_0 \to \mu_b^\prime$ in the TMDs, one obtains the Sudakov factor at this new scale $\mu_b^\prime$ which does not
become larger than $Q$
as $b\to 0$. 
This scale choice leads to a significantly larger ${\cal R}$ at low $Q$. Figure \ref{R-regstudy} shows the result for $m_{\chi_{c0}}=3.4$ GeV to 
$m_{\chi_{b0}}=9.9$ GeV for three cases: 1) the unregulated result with the scale $\mu_b$; 2) the result with
regulated $S_A$; and 3) the result with the scale $\mu_b^\prime$ everywhere. 
The variation in the results by a factor of 2-3 gives an indication of the uncertainty in the results in Fig.\
\ref{RplotQT}. At $Q_T=0$ this uncertainty is not related to the inclusion of the $Y$ term, but rather with higher order
corrections. In addition, there is uncertainty from nonperturbative contributions, which will be discussed in the
following. 

\begin{figure}[htb]
\begin{center}
\includegraphics[height=7 cm]{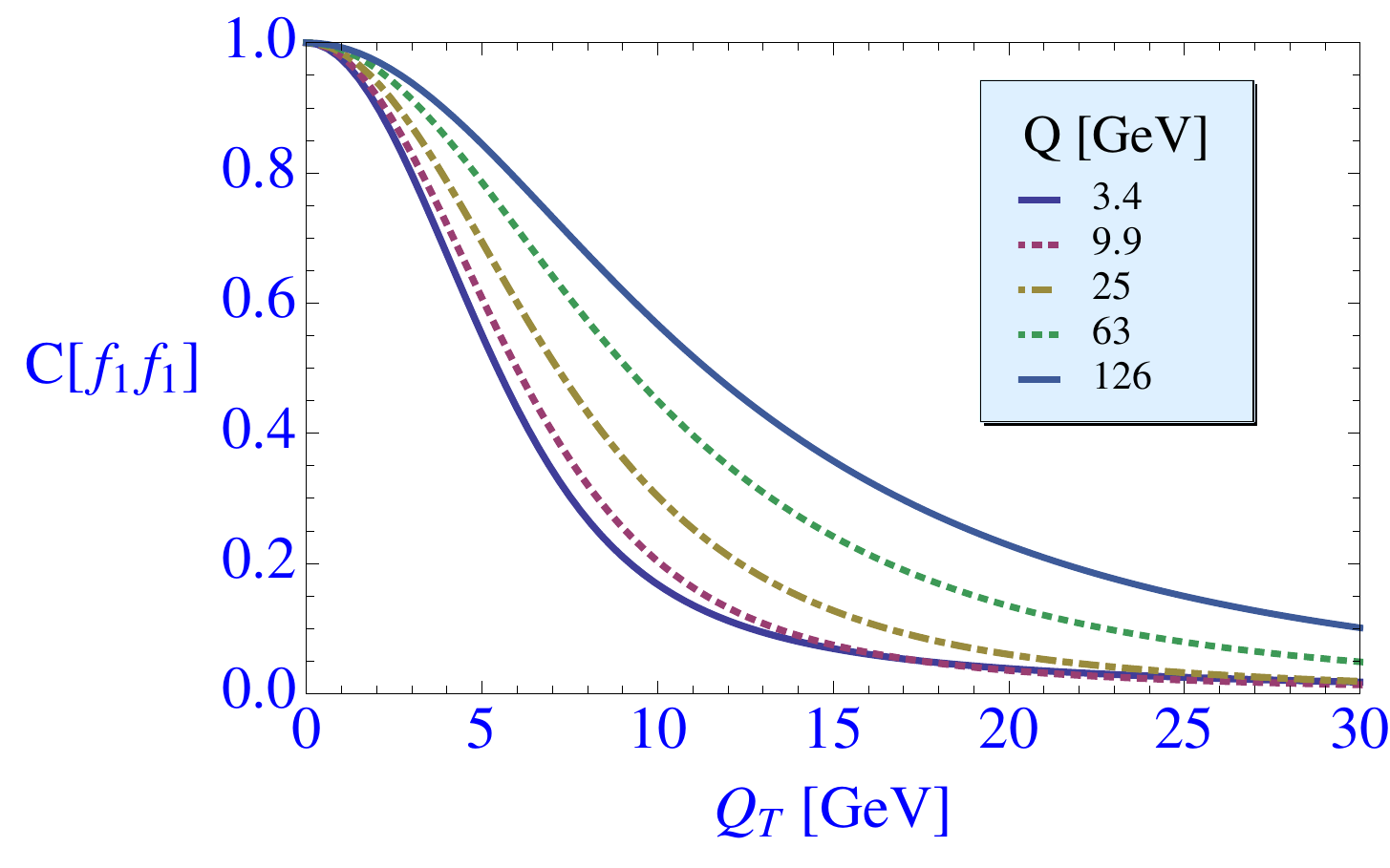}\hspace{1 cm}
\caption{Eq.\ \eqref{convf1}
scaled and plotted for $R=0.5\, {\rm GeV}^{-1}$ and energy scales from $Q=3.4\, \text{GeV}$ to $Q= 126\, \text{GeV}$.}
\label{Gaussgetstail}
\end{center}
\vspace{-2 mm}
\end{figure}

In the above we have only included the perturbative tails, but 
for less small $b$, i.e.\ $b\sim 1/M$, when $b_*$ is close to $b_{\max}=1.5 \, {\rm GeV}^{-1}$, 
one can become sensitive to an intrinsically present nonperturbative contribution, 
often modeled by Gaussians for $Q_T \sim M$ values. 
If one were to include only Gaussians, a completely different conclusion about the $Q$ dependence would be reached. 
Although unrealistic, let us for illustration purposes take for the unpolarized TMD:  
$f_1^g(x,p_T^2)= f_1^g(x) R^2 \exp(-p_{T}^2 R^2)/\pi$, such that $\tilde{f}_1^{g}(x,b^2)=  f_1^g(x) \exp(-b^2/(4 R^2))$ and consequently,  
\beq 
\mathcal{C}\left[f_{1}^{g}\, f_{1}^{g}\right]  =   \frac{1}{2\pi} \int_0^\infty db b J_0(bQ_T^{}) e^{-S_A(b,Q)} e^{-b^2/(2 R^2)} f_1^g(x_A;\mu_b)  
f_1^g(x_B; \mu_b).
\label{convf1}
\eeq
Here we do not employ the $b_*$ method or include $S_{NP}$, 
because the Gaussian will act as a cut-off on $b$ and we do not aim to make it realistic. 
This expression exhibits the general property of TMD 
evolution that a transverse momentum distribution that is approximately Gaussian at some 
low scale $Q \sim M$ will develop a power law tail in transverse momentum at large $Q$ (cf.\ also e.g.\ \cite{AR}). 
This is shown in Fig.\ \ref{Gaussgetstail}, where for illustration purposes we display the curves also for $Q_T$ values beyond $Q/2$. 
In order to avoid negative cross sections at large $Q_T$, the result with $\mu_b \to \mu_b^\prime$ in Eq.\ (\ref{SAmubeval})  is considered.

For the linearly polarized gluon TMD we choose \cite{Boer:2009nc} 
$h_1^{\perp g}(x,p_T^2)= c M^2  f_1^g(x) R_h^4 \exp(-p_{T}^2 R_h^2)/\pi$, because the function does not need to vanish at $p_T =0$. 
Using Eq.\ (\ref{defhtilde}) one finds $\tilde{h}_1^{\perp g}(x,b^2)= c b^2  f_1^g(x) \exp(-b^2/(4 R_h^2))/(8 R_h^2)$, which satisfies $\tilde{h}_1^{\perp g}(x,0)=0$ as expected from $J_2(0)=0$. 
In order to satisfy a Soffer-like bound, one must 
choose $R_h > R$. Just as in Ref.\ \cite{Boer:2009nc} we therefore take $R_h^2=R^2/r$ for some $r <1$ and choose 
$r$ and $c$ such as to maximize\footnote{There are indications 
\cite{Metz:2011wb,Dominguez:2011br} that at small $x$ the Soffer bound is in fact saturated.} $h_1^{\perp g}$, i.e.\ $r=2/3$ and $c=2er(1-r) \approx 1.2$.

The Gaussian functional form results in
\beq
\mathcal{C}\left[\frac{(\bm p_{T}\cdot\bm k_{T})^{2}-\tfrac{1}{2}\bm p_{T}^{2}
\bm k_{T}^{2}}{2M^4}\, h_{1}^{\perp g}\, h_{1}^{\perp g}\right]  
  =  \frac{c^2}{64 R_h^4} \frac{1}{2\pi} \int_0^\infty db\, b^5 J_0(bQ_T^{}) e^{-S_A(b,Q)}\, e^{-b^2/(2 R_h^2)} f_1^g(x_A;\mu_b)  
f_1^g(x_B; \mu_b).
\eeq
This expression indeed exhibits two nodes in $Q_T$.    
The ratio ${\cal R}$ in the Gaussian model at small $Q_T$ thus becomes:
\beq 
{\cal R}(Q_T)= \frac{c^2 \int_0^\infty db b^5 J_0(bQ_T^{}) e^{-S_A(b,Q)}
e^{-b^2/(2 R_h^2)}f_1^g(x_A;\mu_b)  
f_1^g(x_B; \mu_b)}{64 R_h^4 
 \int_0^\infty db b J_0(bQ_T^{}) e^{-S_A(b,Q)} e^{-b^2/(2 R^2)}f_1^g(x_A;\mu_b)  
f_1^g(x_B; \mu_b)}.
 \eeq
Numerically this quantity is very small for all $Q$. It is of order $10^{-3}$ for the choice $R=0.5\, {\rm
GeV}^{-1}$ and $10^{-5}$ for the choice $R=2\, {\rm GeV}^{-1}$, only becoming percent level for very small $R
\simorder 0.2 \, {\rm GeV}^{-1}$. At $Q_T=0$ it falls off with $Q$ roughly as $1/Q^{0.9}$ for both $R=0.5\, {\rm
GeV}^{-1}$ and $R=2\, {\rm GeV}^{-1}$ for $Q \simorder 20\, {\rm GeV}$. This Sudakov suppression is primarily due to the
additional power of $b^4$ in the numerator. This can be illustrated by the following simplified, but analytic analysis.
For $S_A$ in Eq.\ (\ref{SAmub}) (dropping the second, constant term in square brackets) 
one can derive the following analytic result for a ratio that is essentially ${\cal R}(0)$ for $n=4$, but without the
Gaussians and scale dependent TMDs:
\beq
\frac{\int_0^\infty db^2 \, b^n \,  \exp\left({-S_{A}(b,Q)}\right)}{\int_0^\infty db^2 \, 
\exp\left({-S_{A}(b,Q)}\right)} =  c_n \; 
\left(\frac{b_0^2}{\Lambda^2}\right)^{\frac{n}{2}}
\;\left( \frac{Q^2}{\Lambda^2} \right)^{\frac{C_A}{\beta_1} \ln c_n},
\eeq
where $c_n = (1 + C_A/\beta_1)/(1 + n/2 + C_A/\beta_1)$. For $n=4$ and 5 flavors, the scale dependence of this ratio is $Q^{-1.80}$. 
Addition of the Gaussian factors and TMDs makes ${\cal R}$ at $Q_T=0$ fall off more slowly with $Q$, but it is clear
that in the Gaussian model
linear gluon polarization is irrelevant, even at low scales. As said, this is not a realistic model.

In order to study the combined effect of an intrinsically present nonperturbative $h_1^{\perp g}$ and of perturbative tails, 
we consider TMDs that are approximately Gaussian at small transverse momentum, 
but have the proper power law fall-off at large transverse momentum:
\ba
f_1^g(x,p_T^2)& = & f_1^g(x) \frac{R^2}{2\pi} \frac{1}{1+p_T^2 R^2},\nonumber\\
h_1^{\perp g}(x,p_T^2)& = & c f_1^g(x) \frac{M^2 R_h^4}{2\pi} \frac{1}{(1+p_T^2 R_h^2)^2},
\ea
such that 
\ba
\tilde{f}_1^{g}(x,b^2) & = &  f_1^g(x) K_0(b/R), \label{f1besselK}\\
\tilde{h}_1^{\perp g}(x,b^2) & = & \frac{c}{4} f_1^g(x) \frac{b}{R_h} K_1(b/R_h). \label{besselK}
\ea
Note that the last expression holds strictly speaking only for nonzero $b$. In this form it does 
not exhibit the property $\tilde{h}_1^{\perp g}(x,0)=0$.

In comparing $\tilde{f}_1^{g}(x,b^2)$ in Eqs.\ (\ref{f1besselK}) and (\ref{f1tail}) it should be realized 
that the latter is the form that enters in the resummed expression, whereas the former still includes 
the large logarithm of $b$ at small $b$. The expressions appropriate for use in the expression including 
the Sudakov factor one should divide out this logarithm, which for $b\geq b_0$ requires 
regularization. Our Gaussian+tail model therefore is as follows:
\ba
\tilde{f}_1^{g}(x,b^2;\mu_b^2,\mu_b) & = &  f_1^g(x;\mu_b) K_0(b/R)/\ln(R b_0/b+1), \label{f1besselK2}\\
\tilde{h}_1^{\perp g}(x,b^2;\mu_b^2,\mu_b) & = & \frac{c}{4} f_1^g(x;\mu_b) \frac{b}{R_h} K_1(b/R_h)/\ln(R_h b_0/b+1).
\label{besselK2}
\ea
This model expression does exhibit the property $\tilde{h}_1^{\perp g}(x,0)=0$.

For the numerical study we take $r=R^2/R_h^2=2/3$ and $c=2$ which is the maximum value allowed to satisfy the
upper bound $p_T^2 |h_1^{\perp g}(x,p_T^2)| /2M^2 \leq f_1^g(x,p_T^2)$ for {\em all\/} $p_T$ values, 
although $h_1^{\perp g}$ could in principle be substantially larger at smaller $p_T$. 
For $c=2$ the bound is only saturated in the limit $p_T \to \infty$. 
Since the ``width'' $R$ of the distribution is associated with the intrinsic transverse momentum
it seems appropriate to take $R=2\, {\rm GeV}^{-1}$.
We find that the ratio ${\cal R}$ increases for smaller $R$ choices, like in the Gaussian case. 

The ratio $\tilde{h}_1^{\perp g}(x,b^2)/\tilde{f}_1^{g}(x,b^2)$ grows as a function of $b$, but is not identical to
the 
ratio of the tail-only expressions even at small $b$. 
The difference is due to the Fourier transformation that for all nonzero $b$ is sensitive to the small $p_T$ behavior of the TMDs. 
For $b\simorder 1$ the ratio of TMDs for the tail-only case becomes significantly larger than for this model.

\begin{figure}[htb]
\begin{center}
\includegraphics[height=7 cm]{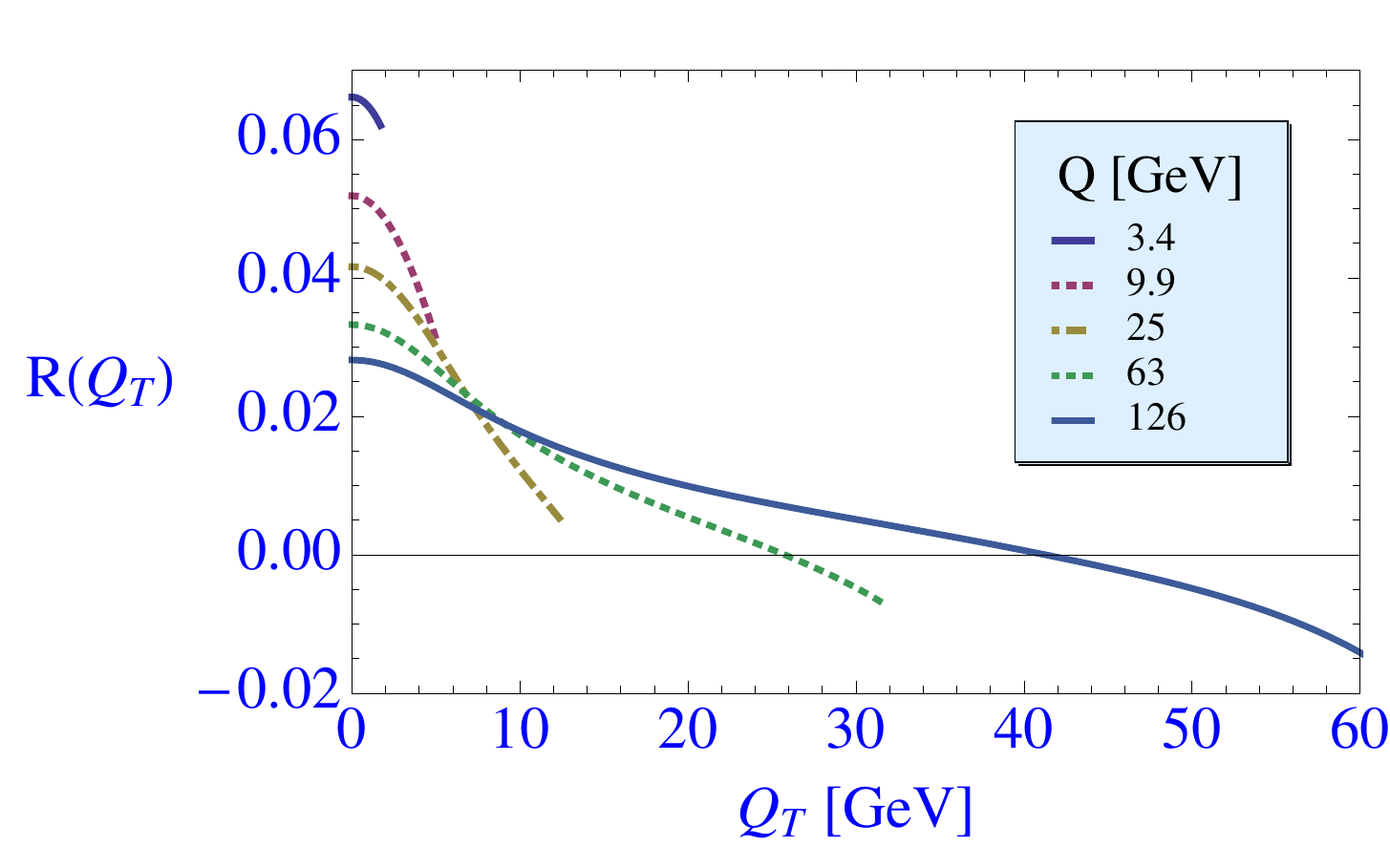}\hspace{1 cm}
\caption{${\cal R}(Q_T)$ evaluated using the Gaussian+tail model in Eqs.\ (\ref{f1besselK2}) and (\ref{besselK2}) 
at $Q=3.4, 9.9, 25, 63, 126 \, \text{GeV}$.}
\label{BesselKmodel}
\end{center}
\vspace{-2 mm}
\end{figure}

Figure \ref{BesselKmodel} shows the model results for ${\cal R}$. 
The results are significantly different from the tail-only results in Fig.\ \ref{RplotQT}. 
Addition of $b_*$ and $S_{NP}$ does not change the results much. 

\begin{figure}[htb]
\begin{center}
\includegraphics[height=7 cm]{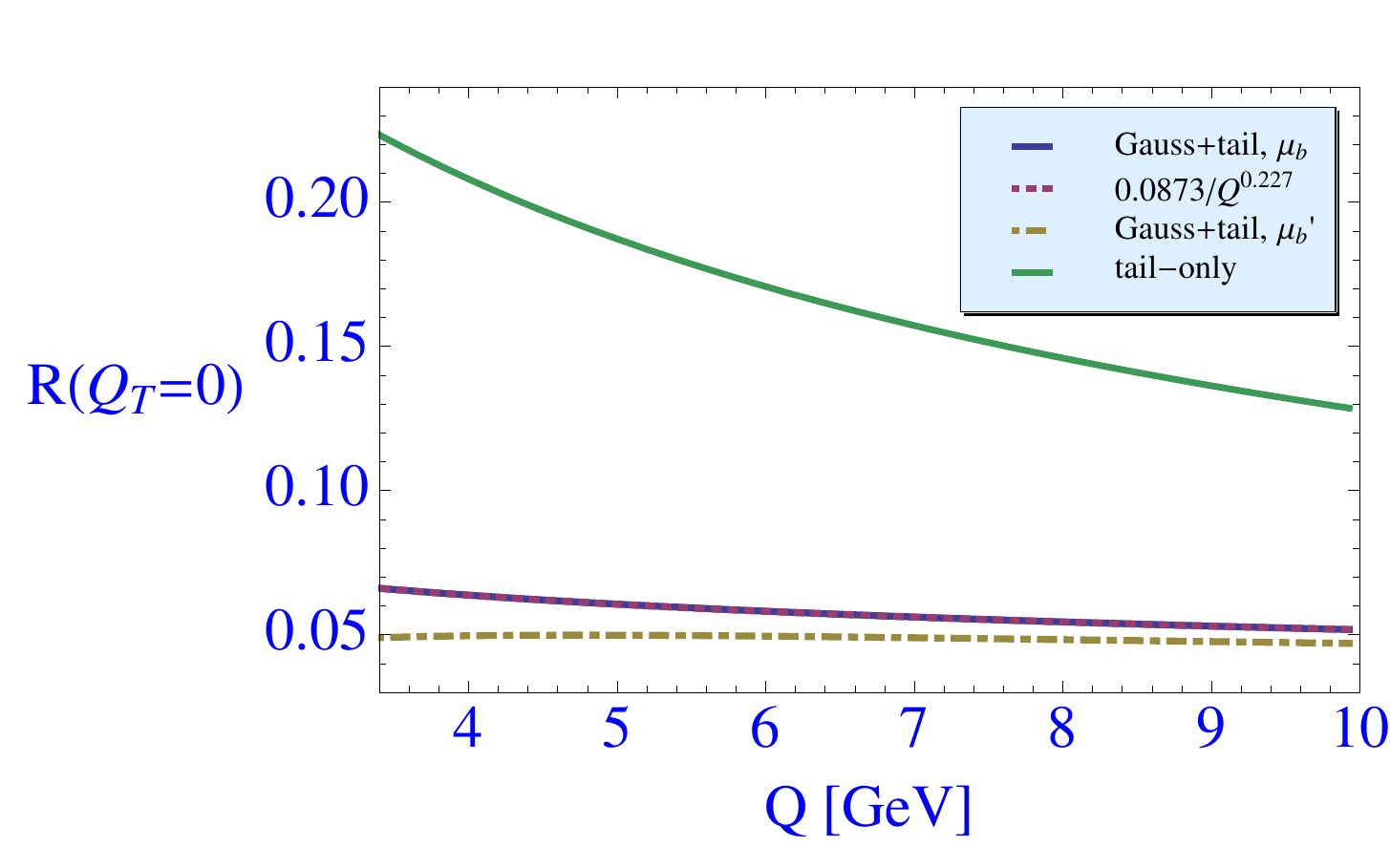}\hspace{1 cm}
\caption{${\cal R}(Q_T)$ at $Q_T=0$ as a function of $Q$ in the region between $m_{\chi_{c0}}=3.4$ GeV
and $m_{\chi_{b0}}=9.9$ GeV for the unregulated ($\mu_b$) and regulated ($\mu_b^\prime$) Gaussian+tail model
and for the tail-only case.}
\label{BesselKmodelQ}
\end{center}
\vspace{-2 mm}
\end{figure}

The above results are obtained without regulator for very small $b$ values. Replacing $\mu_b \to \mu_b^\prime$ does not 
significantly alter the result (less than $5\%$) for $Q\simorder 20$ GeV, where ${\cal R}(Q_T=0)$ to good approximation 
falls off as $1/Q^{0.24}$. Even in the low $Q$ region the effect of choosing the regulator scale $\mu_b^\prime$ is maximally 
25\% in the studied region, as can be seen in Fig.\ \ref{BesselKmodelQ}, which displays ${\cal R}(Q_T=0)$ in the region 
between $m_{\chi_{c0}}=3.4$ GeV and $m_{\chi_{b0}}=9.9$ GeV for the unregulated ($\mu_b$) and regulated 
($\mu_b^\prime$) Gaussian+tail model together with the tail-only case. As can be seen, at $Q_T=0$ the $Q$ dependence of ${\cal R}$ in this region is to very good approximation 
described by a $1/Q^{0.23}$ fall off in the unregulated case, but becomes almost flat in the $\mu_b^\prime$ case, developing 
a slight maximum around $Q=5$ GeV. In all cases there is only a mild evolution in the Gaussian+tail model, as compared to 
the tail-only case.

Although the model may not be fully realistic, it does illustrate the difference that can exist between 
the results obtained from a TMD formalism as compared to a CSS formalism that by construction is not sensitive 
to the specific form of the nonperturbative contributions to the TMDs from small $p_T$.
The CSS formalism includes nonperturbative contributions only through the nonperturbative Sudakov factor $S_{NP}$, 
whereas the TMDs as function of $b$ (and of $b_*$) are also dependent on the nonperturbative region $p_T \sim M$
to some extent. 
The behavior of the TMDs at small $p_T$ can thereby affect the results for all $Q_T$, 
thanks to the Fourier transform receiving contributions from large $b$ values too (i.e.\ when $b_* \sim
b_{\max}$). 

As there is only mild to almost no evolution 
between the two quarkonium states in the Gaussian+tail model, as compared to the tail-only case, the actual amount of TMD evolution  
observed could in principle constrain the nonperturbative contribution and give an expectation for (or be consistent with) the 
magnitude at the Higgs mass scale.

\section{Summary and conclusions}
The TMD evolution of the transverse momentum distribution of a colorless scalar boson produced in proton-proton collisions has been studied. 
The main objective was to get a quantitative estimate of the relative contribution ${\cal R}$ from linearly polarized gluons in Higgs production. Using the perturbatively calculable small-$b$ dependence of the TMDs --the perturbative tails-- and the Sudakov factor, ${\cal R}$ was found to be on the percent level at the Higgs mass scale. This tail-only estimate is a factor 2-3 larger than the earlier estimate of \cite{Wang:2012xs} within the CSS approach, 
which includes some higher order corrections and uses a different $S_{NP}$. In addition, in the TMD approach there can be nonperturbative contributions that go beyond the nonperturbative Sudakov factor of the CSS approach. These nonperturbative contributions unfortunately cannot be calculated.
To investigate their relevance a model was considered in which the TMDs are approximately
Gaussian at small transverse momentum, but exhibit the correct power law fall-off of the perturbative tail at large
transverse momentum. This model also shows percent level effects from linear gluon polarization at the Higgs mass scale, but reached from lower $Q$ values by a considerably slower evolution. The differences between the tail-only results and those from the Gaussian+tail model indicate that the behavior of the TMDs 
at small $p_T$ values can in principle be important for all $Q_T$ and $Q$.
In the Gaussian+tail model there is only modest evolution compared to the tail-only case, therefore, the actual amount of TMD evolution  
observed could in principle constrain the nonperturbative contribution. Just for comparison,  also a pure Gaussian
model was considered, leading for reasonable choices for the width to very small effects, which are well below the percent level even at low
energy scales. In addition, the Gaussian-only model suffers from rather strong Sudakov suppression with increasing
energy scale.

Since the presented calculations apply to colorless scalar boson production of varying mass $Q$, one can also consider
$C$-even scalar quarkonium states $\chi_{c0}$ and $\chi_{b0}$ as discussed in \cite{Boer:2012bt}. The measurement of
${\cal R}$ for those two quarkonium states would allow to check the rather fast evolution from $m_{\chi_{c0}}=3.4$ GeV
to $m_{\chi_{b0}}=9.9$ GeV obtained in the tail-only calculation presented here. On the other hand, in the Gaussian+tail
model much less TMD evolution is observed in this low $Q$ region. This could serve as a means to constrain the
nonperturbative contributions from the TMDs at small transverse momenta, i.e.\ the intrinsic contribution from linearly
polarized gluons. An additional source of uncertainty here arises from the very small-$b$ region. The $b$ region around
and below $1/Q$ is quite important, even for small $Q_T$ and especially at lower $Q$. This affects the estimates for the
low mass quarkonium states which become uncertain within at least a 
factor of 2-3 in the tail-only case. Despite all the uncertainties in the estimates, the results clearly indicate that the 
effects of linearly polarized gluons need not be as small as the CSS result of \cite{Wang:2012xs} suggests and 
hopefully investigations at the LHC can offer experimental information about these effects.

Similar studies for the pseudoscalar case ($\eta_c, \eta_b$) and for angular modulations of the transverse momentum distribution 
can be done in a straightforward manner too. Finally we point out that in the color evaporation picture also states like the $\Upsilon$ 
can be considered, as in \cite{Kulesza:2003wi}, which are more readily measured. However, other arguments suggest that such $J=1$ 
states may not be sensitive to the linear polarization of gluons at small transverse momentum, except perhaps at subleading orders
\cite{Boer:2012bt}. Nevertheless, experimental studies of the unpolarized gluon TMD using quarkonium states are of
interest in their own right as discussed recently in \cite{Dunnen:2014eta}.

\begin{acknowledgments}
The authors thank Miguel Garc\'ia Echevarr\'ia, Tomas Kasemets, Cristian Pisano, and Werner Vogelsang for useful discussions.
This work was supported in part by the German Bundesministerium f\"{u}r Bildung und Forschung (BMBF),
grant no. 05P12VTCTG.
\end{acknowledgments}


\end{document}